\documentclass[runningheads,a4paper]{llncs}

\usepackage[T1]{fontenc}
\usepackage{graphicx}
\usepackage{booktabs}
\usepackage{amsmath}
\usepackage{amssymb}
\usepackage{multirow}
\usepackage{hyperref}

\usepackage[protrusion=true,expansion=false]{microtype}
\begin{document}

\title{Data-Efficient Multimodal Alignment for Histopathology-based Molecular Prediction}
\titlerunning{Molecular Prompting via H\&E-RNA Latent Space Alignment}

 \author{Anonymized Authors}
 \authorrunning{Anonymized Author et al.}
 \institute{Anonymized Affiliations\\
     \email{email@anonymized.com}}
\author{Dominik Winter\inst{1} \and Dominik Vonficht\inst{1} \and Lo\"{i}c Le Bescond\inst{1} \and Christian Gebbe\inst{1} \and Marco Rosati \inst{1} \and Richard J. Chen\inst{1} \and Markus Schick\inst{1} \and \\ Ross Stewart\inst{2} \and Nicolas Brieu\inst{1}}                            
\authorrunning{Winter et al.}                                                                                                    
\institute{AstraZeneca Computational Pathology and Biomarkers, Munich, Germany \and AstraZeneca, Early Oncology and Translation Medicine, Cambridge, UK}    
\maketitle

\begin{abstract}
H\&E-stained whole-slide images offer cohort-scale availability and rich spatial context but lack molecular specificity, 
whereas bulk RNA-seq provides transcriptome-wide resolution at high cost with limited archival availability. 
We show that training a lightweight alignment module atop frozen histopathology and RNA-Seq foundation models enables \emph{open-vocabulary molecular prompting}-querying H\&E slides with gene-set signatures to predict pathway activity without sequencing or end-to-end retraining. 
Using contrastive learning on a multi-cancer cohort ($N{=}1{,}720$), we achieve a 25-fold improvement in retrieval over baseline methods. 
Systematic analysis reveals a graduated predictability spectrum: morphologically grounded programs (cell-cycle programs, immune-related) are most reliably predicted ($R^2{>}0.5$), 
while predicting pathways with no morphological footprint remains challenging as expected. 
We validate clinical utility on the POSEIDON clinical trial: 
H\&E-predicted squamous cell carcinoma scores recapitulate NSCLC subtype identity and predicted IFN-$\gamma$ mirror PD-L1 tumor-cell expression groups. 
Furthermore, genesets describing immune activation and fibrosis predict known tumor microenvironment archetypes from histology alone.
We further validate generalization of our approach across unseen cohorts and demonstrate data-efficient domain adaptation, establishing a slide-native framework for molecular analysis on H\&E images.

\end{abstract}

\section{Introduction}
\label{sec:intro}
\begin{figure}[htb]
  \centering
  \includegraphics[width=0.9\linewidth]{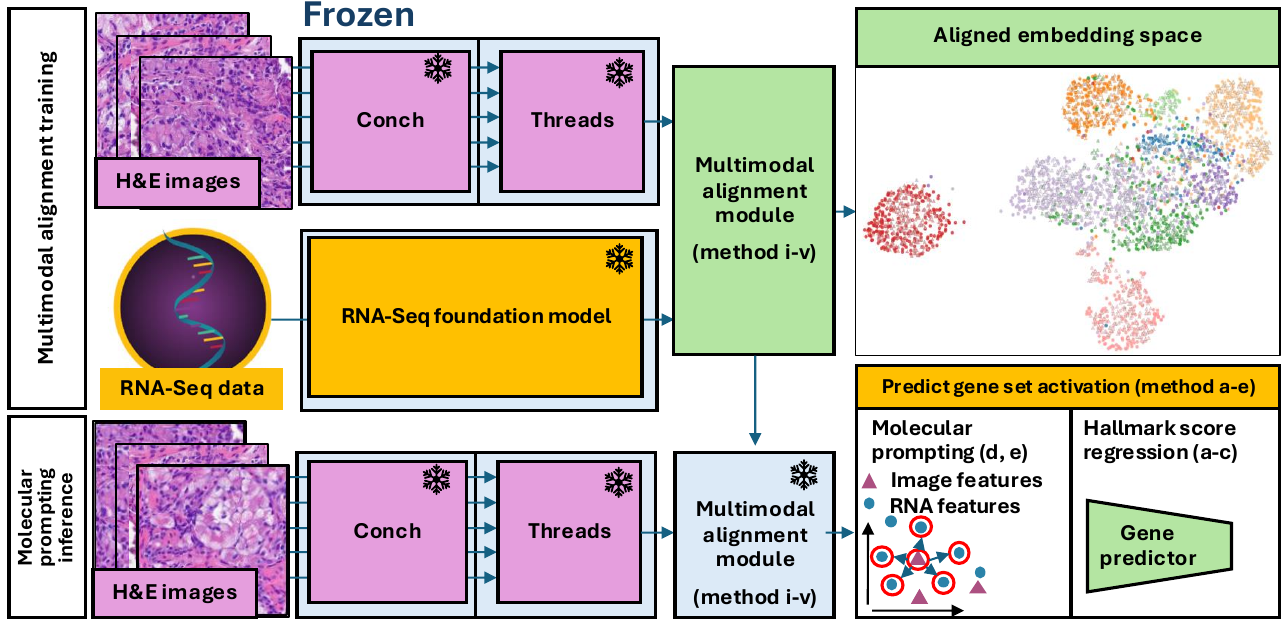}
\caption{\textbf{Overview.} (top) Frozen foundation models and an alignment module (i-v) map H\&E and RNA-Seq to a shared latent space. (bottom) During inference, gene sets act as \emph{open-vocabulary molecular prompts}, allowing H\&E embeddings to predict pathway activity by querying an RNA reference database via Soft-kNN or a trained predictor (a-e).}
  \label{fig:overview}
\end{figure}

H\&E‑stained tissue sections are central to diagnostic pathology.
Self-supervised foundation models trained on millions of digitized WSIs~\cite{chen2024uni,lu2024conch,vaidya2025molecular} enable robust image analysis, such as cancer subtyping, survival prediction, and biomarker discovery ~\cite{vaidya2025molecular,ding2025multimodal,kather2019msi,kather2019predicting}. 
However, H\&E captures morphology rather than direct transcriptomic pathway activity. 
Bulk RNA-seq provides transcriptome-wide resolution but is expensive and is often absent from archival cohorts~\cite{weinstein2013tcga}.
Early works demonstrated that transcriptomic signals and biomarkers, such as microsatellite instability, can be predicted directly from H\&E stained sections~\cite{kather2020pan,schmauch2020he2rna}. 
Recent frameworks have formalized this into supervised pathway regression. TIGER~\cite{howard2025integration} maps embeddings of H\&E stained WSI to breast-cancer-specific gene signatures, while HistoPrism~\cite{hu2026histoprism} introduces a pan-cancer transformer benchmarked on the 50 MSigDB Hallmark gene sets. OmniScreen uses the Virchow image foundation model to predict gene biomarkers~\cite{wang2024screen}. However, these supervised methods require response labels, map exclusively to fixed predefined output spaces, cannot query novel gene sets at inference time, and yield no shared retrieval space.
Foundation models for WSIs~\cite{chen2024uni,lu2024conch,vaidya2025molecular} and transcriptomics~\cite{cui2024scgpt} provide powerful data representations. 
THREADS~\cite{vaidya2025molecular} is trained to generate H\&E WSI representations by aligning genomic and transcriptomic profiles in a shared embedding space via its image encoder and gene encoder (scGPT~\cite{cui2024scgpt} with a custom RNA-head~\cite{vaidya2025molecular}), though it is not primarily designed to ensure exact alignment between image and gene embeddings. 

This leads us to explore in this work further alignment strategies building on the existing THREADS image and gene models for histopathology and bulk transcriptomics at a multi-cancer scale, introducing \emph{open-vocabulary molecular prompting}.
Further improving the alignment of H\&E and RNA-Seq latent spaces beyond the out-of-the-box THREADS models enables the following capabilities: (1)~\emph{cross-modal retrieval} for identifying molecularly similar cases and (2)~\emph{open-vocabulary molecular prompting}—using any gene set from any database as an annotation-free query at inference without modifying the model; which we compare to (3)~\emph{hallmark regression} heads that are trained to predict set gene sets directly from H\&E as done in previous works~\cite{hu2026histoprism,howard2025integration}. 

Thereby we enable retrospective deployment on archival cohorts lacking prospective RNA-seq. Our framework shifts the paradigm from fixed-target prediction to open-vocabulary prompting for \emph{any} arbitrary gene set at inference time.
Our contributions are: (1) introduction of a strategy for alignment of H\&E and RNA-Seq embeddings using existing foundation models, including a systematic benchmark of five H\&E-RNA alignment strategies across two RNA encoders; (2) \emph{open-vocabulary molecular prompting} benchmarked against supervised regression heads; (3) clinical validation on the POSEIDON clinical trial in an inference-only setting; and (4) a data-efficient domain-adaptation across two datasets. 

\section{Method}
\label{sec:method}
We describe a lightweight cross-modal alignment framework using frozen foundation models and an \emph{open-vocabulary molecular prompting} strategy for gene set prediction from H\&E stained WSIs.

\subsubsection{Problem Formulation and Encoders:}
Let $\mathbf{v}_i \in \mathbb{R}^{d_v}$ and $\mathbf{r}_i \in \mathbb{R}^{d_r}$ be the WSI and RNA-Seq embeddings of patient $i$. Given $N$ paired training samples, we learn a shared latent space $\mathbb{R}^D$ where morphologically and molecularly similar samples are proximate. We use THREADS~\cite{vaidya2025molecular} as the image encoder, aggregating CONCH~\cite{lu2024conch} ViT features via attention MIL into $d_v{=}1024$-dim embeddings. Two RNA encoders are evaluated: \textbf{(a)}~scGPT+THREADS RNA-head~\cite{cui2024scgpt,vaidya2025molecular} ($d_r{=}1024$); \textbf{(b)}~BulkFormer~\cite{kang2025large}, purpose-built for bulk RNA-Seq with self-attention over ${\sim}20{,}000$ gene tokens and sentinel masking ($d_r{=}512$). All encoder weights remain frozen; only a lightweight MLP projection head is trained, making the approach tractable even for modest paired cohorts and preserving the pretrained representations of both foundation models.

\subsubsection{Alignment Strategies:}
We benchmark five strategies (Table~\ref{tab:alignment_train}) using two gene foundation models: 
(i) a no-training baseline using the THREADS image and scGPT+THREADS RNA-head model out-of-the-box~\cite{vaidya2025molecular}, 
(ii) Supervised training using a cross-entropy loss to predict the RNA-Seq embeddings from the H\&E embeddings, 
(iii) Ridge Regression, 
(iv) Canonical Correlation Analysis (CCA) ($d_\mathrm{PCA}{=}128$, $n_\mathrm{CCA}{=}50$), 
(v) CLIP-style contrastive learning. 
Both the Supervised (ii) and CLIP (v) methods utilize a two-layer MLP ($\approx 1.3\mathrm{M}$ parameters) with a residual connection, Layer Normalization, and GELU activation to map the frozen embeddings $\mathbf{r}_i$ into $\ell_2$-normalized representations $\mathbf{z}^r_i \in \mathbb{R}^D$ within the shared latent space. 
For the contrastive approach, a label-aware symmetric InfoNCE loss~\cite{radford2021clip} is used, combining the standard CLIP objective with a supervised contrastive term (weight 0.15) that treats same-cancer-subtype samples as additional positives, with the temperature set to $\tau{=}0.07$. Training uses the AdamW optimizer (lr $3{\times}10^{-4}$) with an effective batch size of 256.

\subsubsection{Molecular Prompting:}
Given a reference RNA library $\{\hat{\mathbf{z}}^r_j\}$ with hallmark scores $\{y^h_j\}$, predicted pathway activity for query image $\hat{\mathbf{z}}^v_i$ is:
\begin{equation}
  \hat{y}^h_i = \sum_{j \in \mathcal{N}_k(i)} w_{ij} y^h_j, \quad
  w_{ij} = \frac{1/d_{ij}}{\sum_{l \in \mathcal{N}_k(i)} 1/d_{il}},
  \label{eq:molprompt}
\end{equation}
where $d_{ij} = 1 - \hat{\mathbf{z}}^v_i \cdot \hat{\mathbf{z}}^r_j$ is the cosine distance between the $\ell_2$-normalized query image embedding $\hat{\mathbf{z}}^v_i$ and reference RNA embedding $\hat{\mathbf{z}}^r_j$, and $\mathcal{N}_k(i)$ denotes the $k{=}5$ nearest RNA-Seq neighbors.
No subtype labels are needed. 
We also evaluate a \emph{Soft-kNN} variant that replaces sparse $k$-NN aggregation with soft-attention weights $w_{ij} \propto \exp(-d_{ij}/\tau)$ computed over the \emph{full} reference library ($\tau{=}0.1$), yielding a smoother, globally-normalised prediction. 

For calculating the \textbf{gene set activity scores}, we use TPM-normalized counts and $\log_2(x{+}1)$-transformed before ssGSEA~\cite{barbie2009ssgsea} scoring on the 50 MSigDB Hallmark gene sets~\cite{liberzon2015msigdb}, yielding $\mathbf{y}_i \in \mathbb{R}^{50}$ per sample, but can be applied to any gene set collection. 
These scores serve as a groundtruth for both supervised regression and molecular prompting benchmarks. Per-hallmark regression heads (RidgeCV, Random Forest, MLP, Soft-kNN) are trained on aligned features via 5-fold cross-validation using the BulkFormer-based CLIP method. 
Domain adaptation fine-tunes BulkFormer CLIP on fractions $\rho \in \{5\%,10\%,25\%,50\%,100\%\}$ of paired target-domain samples.

For clinical validation in an inference-only setting, we use the Soft-kNN variant over the full training-cohort RNA library, with ssGSEA targets normalised within the training cohort only so that no target-domain samples influence the reference scores.

\section{Experiments}
\label{sec:experiments}
We first benchmark the alignment strategies and the molecular prompting methods on a multi-cancer dataset. We then validate our method clinically on the POSEIDON trial and finally ablate domain adaptation on two external cohorts.

\subsubsection{Datasets:}
Models are trained on a multi-cancer dataset ($N{=}1{,}720$) spanning Non-small Cell Lung Cancer (NSCLC) ($n{=}245$), Triple Negative Breast Cancer (TNBC) ($n{=}112$), Renal Cell Carcinoma (RCC) ($n{=}195$), Colorectal Cancer (CRC) ($n{=}187$), ovarian ($n{=}216$), endometrial ($n{=}400$), pancreatic ($n{=}42$), prostate ($n{=}232$), and bladder ($n{=}92$) cancer patient samples. All cross-validation folds are partitioned at the patient level. 
We use the following datasets for evaluating our approach in several experiments: \textbf{POSEIDON} ($n{=}265$, NSCLC ~\cite{johnson2023poseidon}), \textbf{TCGA-BRCA} ($n{=}1{,}042$, pan-breast), and \textbf{TCGA-LUAD} ($n{=}223$, lung adenocarcinoma).

\subsubsection{Alignment Benchmark:}
BulkFormer MLP-CLIP achieves the best performance in R@5${=}56.0\%$ and R@10${=}71.9\%$ with competitive scoring in cosine similarity and Maximum Mean Discrepancy (MMD) (Table~\ref{tab:alignment_train}) in a 5-fold cross-validation experiment. 
The no-training baseline and supervised training yield near-zero retrieval performance.
BulkFormer MLP-CLIP consistently outperforms scGPT pairings, making it our method of choice for further experiments.

\begin{table}[tb]
\caption{Alignment quality on multi-cancer dataset (5-fold cross-validation), img$\to$rna direction. Recall@$K$ (\%) as mean$\pm$std. \textbf{Bold}: best per metric.}
\label{tab:alignment_train}
\centering
\resizebox{\linewidth}{!}{
\begin{tabular}{@{}llccccc@{}}
\toprule
Method & RNA encoder
  & R@1$\uparrow$ & R@5$\uparrow$ & R@10$\uparrow$ & Cosine$\uparrow$ & MMD$\downarrow$ \\
\midrule
(i) No Training & scGPT      & $0.3{\pm}0.2$ & $1.4{\pm}0.3$   & $2.8{\pm}0.8$   & $-0.013$ & $0.655$ \\
(ii) Supervised  & scGPT      & $0.1{\pm}0.1$ & $1.2{\pm}0.3$   & $2.7{\pm}0.7$   & $0.001$  & $0.646$ \\
(iii) Ridge       & scGPT      & $8.3{\pm}6.6$ & $27.7{\pm}20.0$ & $38.6{\pm}26.4$ & $0.524$  & $0.276$ \\
(iv) CCA         & scGPT      & $13.5{\pm}1.0$ & $39.2{\pm}3.0$ & $54.5{\pm}1.3$  & $0.315$  & $0.048$ \\
(v) MLP-CLIP        & scGPT      & $11.3{\pm}1.6$ & $38.5{\pm}1.0$ & $55.9{\pm}1.1$  & $0.401$  & $0.121$ \\
\midrule
(ii) Supervised  & BulkFormer & $0.5{\pm}0.3$ & $1.6{\pm}0.5$   & $2.9{\pm}1.0$   & $-0.003$ & $0.611$ \\
(iii) Ridge       & BulkFormer & $16.8{\pm}1.1$ & $45.3{\pm}1.5$ & $62.0{\pm}1.1$  & $\mathbf{0.673}$ & $0.165$ \\
(iv) CCA         & BulkFormer & $\mathbf{23.0{\pm}1.9}$ & $53.0{\pm}0.8$ & $67.8{\pm}2.3$ & $0.406$ & $\mathbf{0.044}$ \\
(v) MLP-CLIP        & BulkFormer & $20.8{\pm}1.5$ & $\mathbf{56.0{\pm}1.2}$ & $\mathbf{71.9{\pm}1.0}$ & $0.501$ & $0.148$ \\
\bottomrule
\end{tabular}}
\end{table}

\subsubsection{Molecular Prompting:}
\label{sec:mol_prompting}
Using BulkFormer with MLP-CLIP, we then benchmark multiple methods for predicting the gene set activation scores, namely direct regression heads (a-c) on the aligned latent space (5-fold CV; $\sigma$ is cross-hallmark std across folds), (d) kNN ($k{=}5$) and (e) Soft-kNN and (Table~\ref{tab:hallmark_prediction}).
Although RidgeCV outperforms Soft-kNN in mean $R^2$, RidgeCV requires a fixed set of labelled hallmark targets at training time and must be retrained for any novel gene set. 
Soft-kNN, by contrast, operates as a true open-vocabulary predictor: any gene set can be queried at inference by computing ssGSEA scores on the reference RNA library, with no model updates required, leading us to use Soft-kNN as method of choice in further experiments.

\subsubsection{Biological Ceiling:}
\label{sec:bio_ceiling}
\begin{figure}[t]
  \centering
  \includegraphics[width=\linewidth]{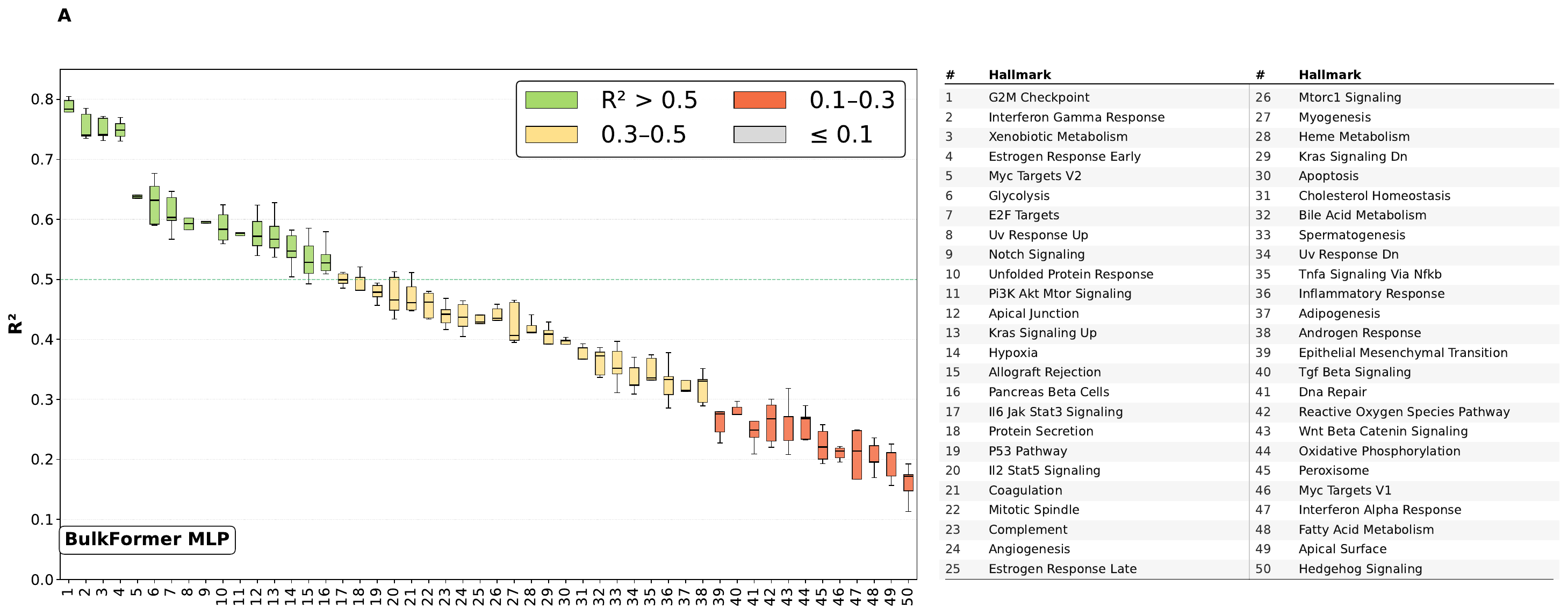}
  \caption{Molecular prompting: predicted vs.\ true ssGSEA scores for MSigDB Hallmark gene sets (multi-cancer, 5-fold cross-validation). Sorted by BulkFormer MLP $R^2$; background shading indicates morphological grounding.}
  \label{fig:hallmarks}
\end{figure}

The per-hallmark $R^2$ pattern on the multi-cancer dataset (5-fold cross-validation) establishes an empirical biological ceiling: gene sets with a morphological footprint in H\&E stained WSIs are predictable, while others are not (Fig.~\ref{fig:hallmarks}), consistent with prior work showing that transcriptomic programs can be predicted from H\&E stained WSIs~\cite{kather2020pan,hoang2024deeppt,pizurica2024sequoia}.

Among \textbf{higher-predictable hallmarks}, cell-cycle programs lead the way: \textsc{G2M Checkpoint} ((1) $R^2{=}0.78$) and \textsc{E2F Targets} ((7) $R^2{=}0.61$), correspond to mitotic figures and nuclear pleomorphism. 
\textsc{IFN-$\gamma$ Response} ((2) $R^2{=}0.75$), \textsc{Allograft Rejection} ((15) $R^2{=}0.53$) and \textsc{(17) IL6-JAK-STAT3 Signaling} ($R^2{=}0.50$) are Immune-related hallmarks which reflect TIL density. 
\textsc{Glycolysis} ((6) $R^2{=}0.63$) and \textsc{Hypoxia} ((14) $R^2{=}0.53$) track necrosis and high-grade architecture. 

\textbf{Lower-predictability hallmarks}, such as \textsc{Oxidative Phosphorylation} ((44) $R^2{=}0.22$), \textsc{Fatty Acid Metabolism} ((48) $R^2{=}0.20$), and \textsc{Androgen Response} ((38) $R^2{=}0.27$) could reflect metabolic pathways with no detectable signal in H\&E stained WSI~\cite{schmauch2020he2rna,kather2020pan,fu2020pan}.

\begin{table}[tb]
\caption{Hallmark gene set prediction performance (5-fold cross-validation). Mean $R^2 \pm$ std across 50 hallmarks, with counts of hallmarks exceeding $R^2 {>} 0.3$ and $R^2 {>} 0.5$.}
\label{tab:hallmark_prediction}
\centering
\begin{tabular}{@{}lccc@{}}
\toprule
Method & Mean $R^2$ $\uparrow$ & Hallmarks ${>}0.3$ $\uparrow$ & Hallmarks ${>}0.5$ $\uparrow$ \\
\midrule
(a) RidgeCV      & $\mathbf{0.465{\pm}0.135}$ & $\mathbf{45/50}$ & $\mathbf{18/50}$ \\
(b) Random Forest & $0.441{\pm}0.132$         & $44/50$          & $16/50$          \\
(c) MLP          & $0.392{\pm}0.148$          & $36/50$          & $15/50$          \\
(d) kNN          & $0.420{\pm}0.130$          & $42/50$          & $16/50$          \\
(e) Soft-kNN     & $0.421{\pm}0.120$          & $44/50$          & $15/50$          \\
\bottomrule
\end{tabular}
\end{table}

\subsubsection{Clinical Validation:}
\label{sec:poseidon}
We validate molecular prompting on POSEIDON~\cite{johnson2023poseidon} ($n{=}265$, NSCLC) in an inference-only experiment.
The MLP-CLIP model is frozen after being trained on the multi-cancer dataset and the Soft-kNN queries a frozen RNA-Seq library  exclusively built from other cohorts (multi-cancer dataset\,$+$\,TCGA-LUAD\,$+$\,TCGA-BRCA). 

\begin{figure}[t]
  \centering
  \includegraphics[width=0.97\linewidth]{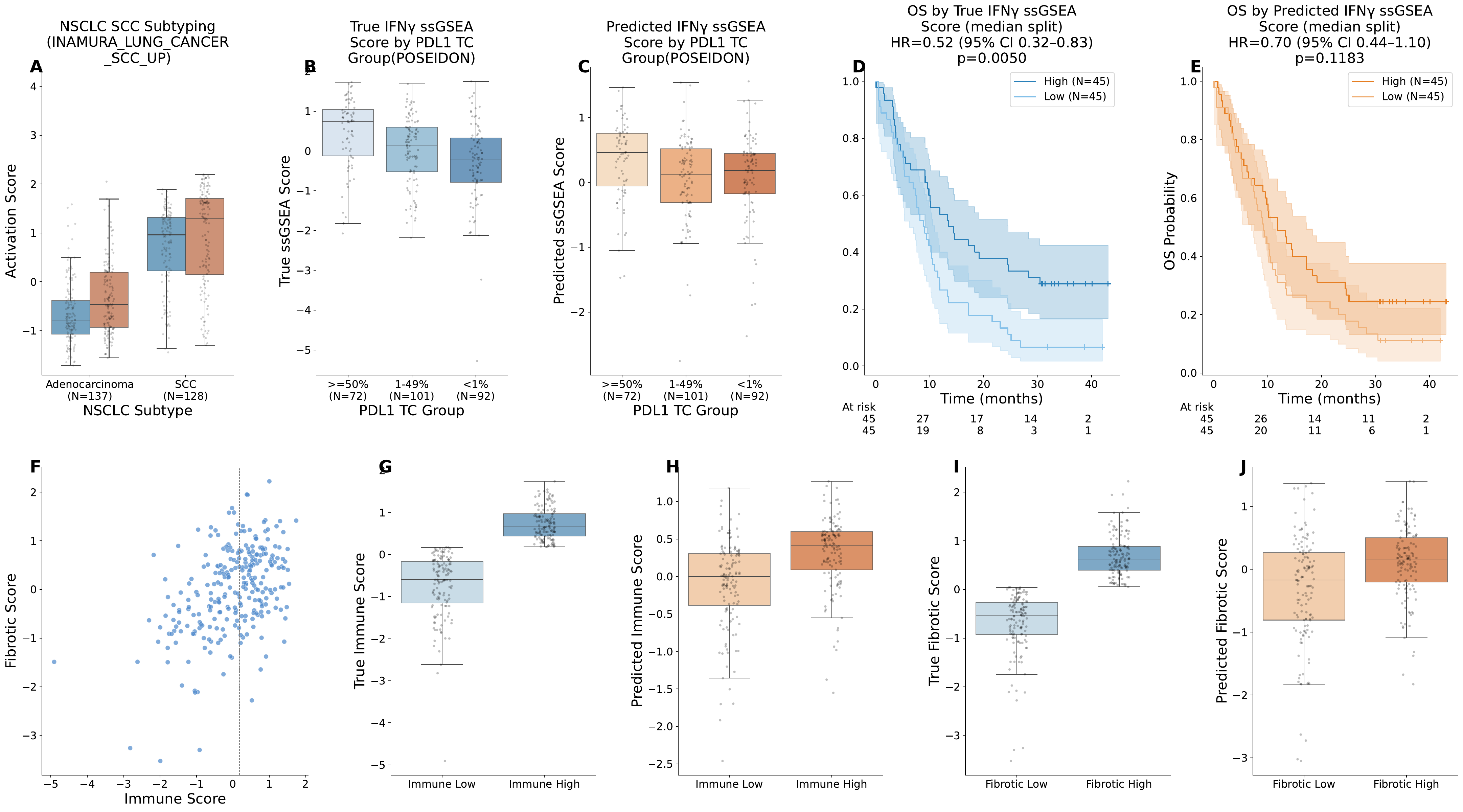}
  \caption{
    \textbf{Clinical validation on POSEIDON:} 
    \textbf{A:}~H\&E-based NSCLC subtyping. 
    \textbf{B-C:}~True and predicted IFN-$\gamma$ ssGSEA vs.\ PD-L1 TC group (all 265 QC-passed patients). 
    \textbf{D-E:}~Kaplan-Meier OS by true/predicted IFN-$\gamma$ median split ($n{=}90$, durvalumab arm). 
    \textbf{F:}~Scatter plot describing true immune (x-axis) and fibrotic (y-axis) scores and classes; 
    \textbf{G-H:}~Boxplots showing true and predicted scores split at true median values into low and high groups. 
    The predicted immune score (H) mirrors the true score (G).
    \textbf{I-J:}~The predicted fibrotic score (J) mirrors the true fibrotic score (I). 
  }
  \label{fig:clinical_validation}
\end{figure}

Using the fixed multimodal aligned space, we predict biologically relevant gene set activity. 
First, we observe that H\&E-predicted Inamura SCC scores separate LUAD from LUSC patients (Fig.~\ref{fig:clinical_validation}A), recapitulating true subtype identity.

Second, we examine IFN-$\gamma$ gene set activity across PD-L1 expression strata:
H\&E-predicted IFN-$\gamma$ scores tend to mirror PD-L1 Tumor Cell (TC) expression groups (Fig.~\ref{fig:clinical_validation}B-C), suggesting H\&E-inferred immune activation may serve as a surrogate for this predictive checkpoint inhibitor biomarker. 
The predicted scores yielded a non-significant but directionally consistent OS stratification in the durvalumab arm (HR=0.70, p=0.118) (Fig.~\ref{fig:clinical_validation}D-E)

Third, to characterize the tumor microenvironment (TME), we assessed molecular programs associated with immune activation and fibrosis, two major determinants of conserved TME archetypes (Fig.~\ref{fig:clinical_validation}F) \cite{bagaev2021}.
We define the immune axis as the mean of the IFN-$\gamma$ and inflammatory response enrichment scores, and the fibrotic axis as the mean of the TGF-$\beta$ and epithelial-mesenchymal transition enrichment scores. 
H\&E-predicted immune and fibrotic scores broadly recapitulated the median-split separation defined by corresponding RNA-seq-derived scores (Fig.~\ref{fig:clinical_validation}G-J).

\subsubsection{Data-Efficient Domain Adaptation}
\begin{figure}[t]
  \centering
  \includegraphics[width=0.97\linewidth]{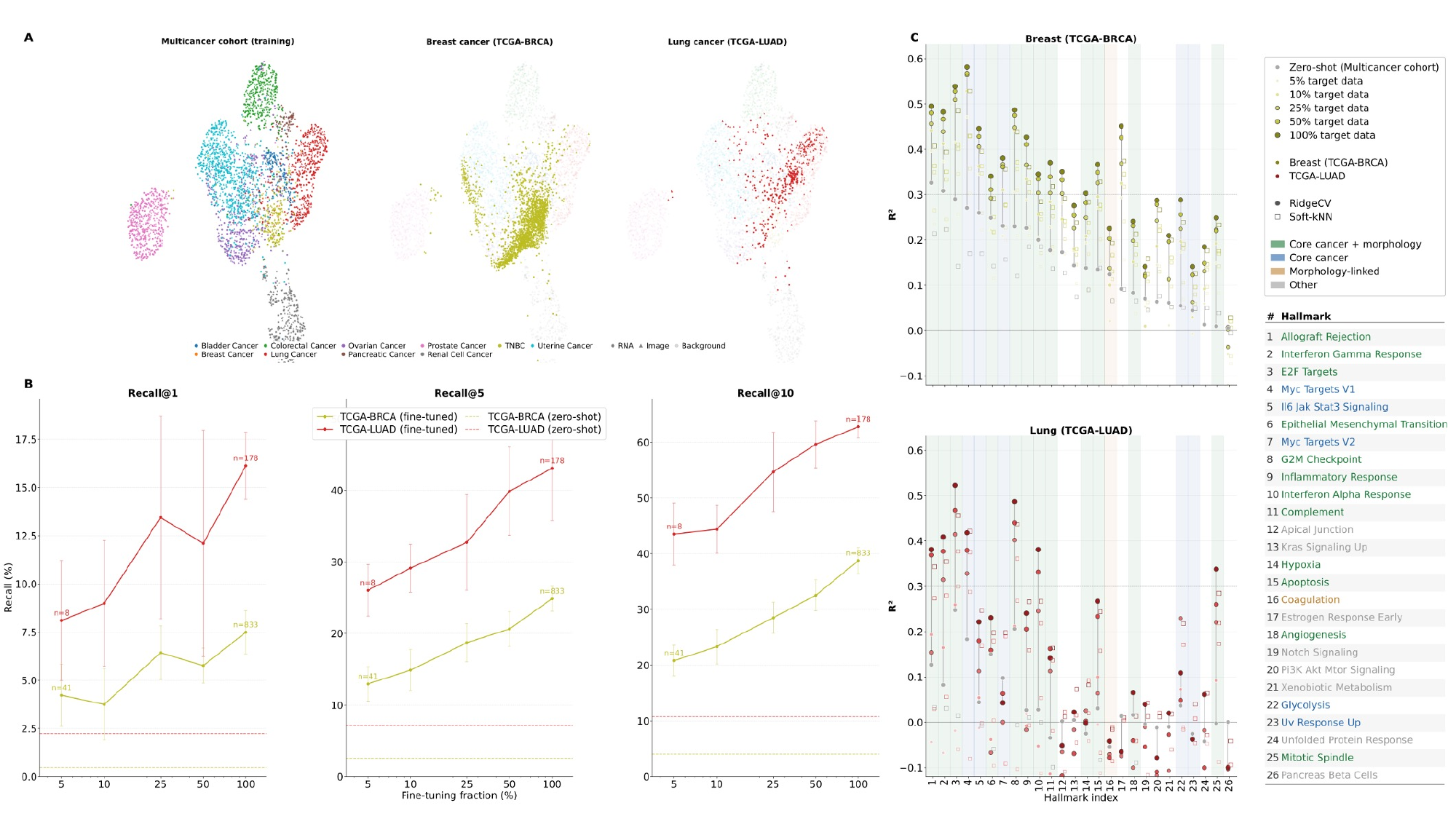}
  \caption{\textbf{Domain adaptation.} \textbf{(A)} UMAPs of the multi-cancer dataset, TCGA-BRCA and TCGA-LUAD embeddings with color-coded indications. \textbf{(B)} Recall@$K$ vs.\ fine-tuning fraction of TCGA-BRCA and TCGA-LUAD datasets. \textbf{(C)} Per-hallmark $R^2$ per fraction; shading groups hallmarks by morphological grounding.}
  \label{fig:combined_finetune}
\end{figure}

Using the multi-cancer trained alignment models on unseen cohorts (TCGA-LUAD, and TCGA-BRCA) exposes a transfer gap: R@10 drops to $10.8\%$, and $4.0\%$, respectively. 
Data-efficient domain adaptation bridges this gap with minimal paired target-domain samples.
Fig.~\ref{fig:combined_finetune}-B shows target-domain fine-tuning with $\rho \in \{5\%,10\%,25\%,50\%,100\%\}$. 
Just 8 paired samples ($5\%$ of TCGA-LUAD) raises R@10 from $10.7\%{\to}43.5\%$, confirming the pretrained model encodes a dense structural prior requiring only minimal target calibration on unseen image and RNA-Seq embeddings. 
Full fine-tuning restores R@10 to $62.8\%$ (TCGA-LUAD), and $38.8\%$ (TCGA-BRCA). 
Hallmark regression (Fig.~\ref{fig:combined_finetune}-C) works best for programs with clear morphological signatures - such as cell cycle, immune activity, and glycolysis - while metabolic pathways remain harder to predict. 
This likely reflects both their weak visual footprint and the complexity of gene sets that span multiple independent processes, making it difficult to find a single consistent morphological correlate. 
The TCGA-LUAD cohort recovers retrieval faster than TCGA-BRCA, which may reflect a greater morphological and molecular heterogeneity.
Please note that current patient data compliance policy associated to the POSEIDON clinical dataset does not allow experiments beyond the 'inference-only' setting presented in section \ref{sec:poseidon}, currently preventing similar fine-tuning experiments to be conducted on this dataset. 

\section{Conclusion}
\label{sec:conclusion}
Training lightweight projection heads atop frozen foundation models enables \emph{open-vocabulary molecular prompting} from routine H\&E slides and querying gene set at inference without sequencing or end-to-end retraining. 
Our systematic benchmark of five alignment strategies establishes BulkFormer MLP-CLIP in combination with Soft-kNN querying as the best-performing method ($71.9\%$ R@10, $25{\times}$ above baseline). 
Per-hallmark analysis reveals a principled predictability spectrum: morphologically grounded programs such as cell-cycle, immune infiltration, and glycolysis tend to be recoverable ($R^2{>}0.5$), while metabolic pathways lacking a tissue-architectural footprint remain challenging.  

Application in an inference-only setting on a clinical trial suggests that frozen molecular prompting can enable biomarker-relevant molecular inference from routine H\&E images at minimal incremental cost.
H\&E-predicted signatures recapitulated established immune and fibrotic archetypes, while the first showed directionally consistent overall survival stratification relative to sequencing-based estimates.
Under domain shift, a few paired samples suffice to bridge the transfer gap, confirming a strong structural prior requiring only minimal target calibration. 
Notable limitations include our modest training cohort size ($N{=}1{,}720$). Future work will explore larger cohorts, additional cancer types, and further clinical trial applications.

\subsubsection*{Acknowledgements:}
All authors are employees of AstraZeneca and some have AstraZeneca shares. 
No author has other relevant financial or non-financial interests to disclose.
We thank the TICA team at AstraZeneca for dataset access.

\bibliographystyle{splncs04}
\bibliography{miccai_paper}

\end{document}